\documentclass[11pt]{article}
\begin{document}

\centerline{\bf\Large Mathematical investigation of the Boltzmann}
 \vskip 5pt\centerline{\bf\Large collisional operator} \vskip 5pt
\centerline{C.Y.~Chen}   

\noindent Dept. of Physics, Beijing University of Aero. and
Astro., Beijing 100083,  PRChina, Email: cychen@buaa.edu.cn

\vskip 5pt
\noindent PACS {51.10.+y} {Kinetic and transport theory of gases}

\begin{abstract}
Problems associated with the Boltzmann collisional operator are unveiled
and discussed. By careful investigation it is shown that
collective effects of molecular collisions in the six-dimensional
position and velocity space are more sophisticated than they appear
to be.
\end{abstract}

The Boltzmann equation was strongly criticized by
Boltzmann's contemporaries and successors. As a subject of serious
debate it involved a large number of scientists and philosophers. One of
the main reasons for having such debate lay in the fact that
Boltzmann
explicitly employed the time reversibility of Newton's law to derive his
kinetic equation while the equation itself appeared to be
time-irreversible. Even today, related paradoxes still bother some of us
and stimulate serious studies (for instance those in chaos theory).

In this brief paper, we wish to report on our recent investigation of
the Boltzmann collisional operator. It is shown that
collective effects of molecular collisions in the six-dimensional
phase space are more sophisticated than they appear to be.

Let's recall key points in the derivation of the Boltzmann
collisional operator given by textbooks of statistical mechanics
\cite{ref:reif}.
For purposes of this paper, only collisions between identical, but still
distinguishable, molecules are considered here. (Such consideration makes
 sense in classical mechanics, not in quantum mechanics.)
 Suppose that a beam of molecules with the initial velocity ${\bf v}_1$
  collide with a molecule
with the initial velocity ${\bf v}_2$.  The scattering cross
section $\bar\sigma$ in the laboratory frame is, according to the standard
 theory, defined in such a way that
\begin{equation}\label{sig} N=\bar\sigma({\bf v}_1,{\bf v}_2\rightarrow
{\bf v}_1^\prime,{\bf v}_2^\prime)
d{\bf v}_1^\prime d{\bf v}_2^\prime \end{equation}
represents the number of type 1 molecules per unit time, per unit flux
emerging after collisions between ${\bf v}_1^\prime$ and
${\bf v}_1^\prime+d{\bf v}_1^\prime$ while the type 2 molecule emerging
between ${\bf v}_2^\prime$ and
${\bf v}_2^\prime+d{\bf v}_2^\prime$.
For exactly the same situation, the scattering cross section in the
center-of-mass frame is defined as such that
$N=\sigma(\Omega)d\Omega$
represents the number of type 1 molecules per unit time
emerging after scattering within the range $d\Omega$ where $\Omega$ is
 the solid angle between ${\bf u} ={\bf v}_2-{\bf v}_1$ and
  ${\bf u}^\prime={\bf v}_2^\prime-{\bf v}_1^\prime$.
The two cross sections are related to each other by
\begin{equation} \label{twosec}  \int_{\Omega}
\sigma(\Omega)d\Omega =\int_{{\bf v}_1^\prime}\int_{{\bf v}_2^\prime}
\bar\sigma({\bf v}_1,{\bf v}_2\rightarrow {\bf v}_1^\prime,{\bf v}_2^\prime)
 d{\bf v}_1^\prime d{\bf v}_2^\prime. \end{equation}
It is assumed in the standard theory that there is time-reversibility of
 collision expressed by
\begin{equation} \label{rev}\bar\sigma({\bf v}_1,{\bf v}_2\rightarrow
{\bf v}_1^\prime,{\bf v}_2^\prime)=\bar\sigma({\bf v}_1^\prime,
{\bf v}_2^\prime\rightarrow {\bf v}_1,{\bf v}_2).\end{equation}
By making use of (\ref{rev}) and (\ref{twosec}),
the net increase of molecules in
a given volume $d{\bf r} d{\bf v}_1$ within the time interval $dt$
is obtained as\cite{ref:reif}\cite{ref:harris}
\begin{equation}\label{cl} dtd{\bf r} d{\bf v}_1 \int_{{\bf v}_2,\Omega}
[f({\bf v}_1^\prime)f({\bf v}_2^\prime)-f({\bf v}_1)f({\bf v}_2)]u\sigma
(\Omega)d\Omega d{\bf v}_2. \end{equation}
In the standard collisionless theory, the net increase per unit time and per
 unit phase volume is found to be $df/dt=0$. By assuming (\ref{cl}) to be a
correction term, we arrive at the Boltzmann equation
\begin{equation}\label{bl} \frac{\partial f}{\partial t}+{\bf v}_1\cdot
\frac{\partial f}{\partial {\bf r}} +\frac
{{\bf F}}m \cdot\frac{\partial f}{\partial {\bf v}_1}=\int_{{\bf
v}_2,\Omega}
[f({\bf v}_1^\prime)f({\bf v}_2^\prime)-f({\bf v}_1)f({\bf v}_2)]u\sigma
(\Omega)d\Omega d{\bf v}_2. \end{equation}
Although the derivation outlined above seems quite stringent, there exist
hidden loopholes. Interestingly enough, even a simple
comparison between the left side and right side of the above
Boltzmann equation offers us delicate things to think about.
On the left side, there is a symmetry between $\bf r$ and ${\bf v}_1$ (in
terms of differentiations). On the right side, the position vector $\bf r$
 serves as an inactive `parameter' and all the operations are
performed in the velocity space. Along this line, many mathematical
 difficulties manifest themselves.

Our first concrete concern is associated with the scattering
cross section in the laboratory frame, namely $\bar\sigma$ defined by
(\ref{sig}). Reconsider the situation in which type 1 molecules (with
${\bf v}_1$ and ${\bf v}_1^\prime$) collide with a type 2 molecule
(with ${\bf v}_2$ and ${\bf v}_2^\prime$). The
conservation laws of energy and momenta imply that
${\bf v}_1^\prime$ and ${\bf v}_2^\prime$ obey, for
any given ${\bf v}_1$ and ${\bf v}_2$,
\begin{equation}
|{\bf v}_2^\prime-{\bf v}_1^\prime|=|{\bf v}_2-{\bf v}_1|=|{\bf u}|
\quad{\rm and}\quad {\bf v}_1^\prime+{\bf v}_2^\prime=
{\bf v}_1+{\bf v}_2=2{\bf c}. \end{equation}
This simply means that all scattered molecules will fall
on a spherical surface with the diameter $|{\bf u}|$ in the
velocity space, called the
accessible surface hereafter. By investigating what takes place in the
velocity space thoroughly, two misconcepts associated with the definition
of $\bar \sigma$ can readily be found.
One is that after $d{\bf v}_1^\prime$ is specified, specifying
 $d{\bf v}_2^\prime$ in
 the definition is a work overdone. The other is that
the cross section should be defined with respect to an area element on
the accessible surface (infinitely thin) rather than with respect to a
volume element (like $d{\bf v}_1^\prime$ in the definition).
The second misconcept is particularly serious in the following sense.
If we assume the volume element $d{\bf v}_1^\prime$ in (\ref{sig}) to
be a small spherical ball whose center lies on the accessible surface,
as shown in Fig.~1a,
the number of molecules entering the small spherical ball per unit time
can be expressed by $N=\rho \pi r^2$,
where $r$ is the radius of the small ball and $\rho$ is the surface
density of melecules on the area
of the accesible surface per unit flux of type 1
molecules. By noting that $\rho$ is finite, we find that
the cross section in (\ref{sig}) is equal to,
with $d{\bf v}_2^\prime$ neglected,
\begin{equation}\label{box}
\bar\sigma=\frac{N} {d{\bf v}_1^\prime}= \frac{\rho\pi r^2}{4\pi
r^3/3}=\frac {3\rho}{4r}, \end{equation} and it tends to infinity
as $r\rightarrow 0$. In contrast with that, if the volume element
$d{\bf v}_1^\prime$ is assumed to be one like a tall-and-slim
cylindrical box shown in Fig.~1b, $\bar\sigma$ tends to zero.
Finally, if $d{\bf v}_1^\prime$ is one shown in Fig.~1c,
$\bar\sigma$ tends to infinity again. All these mean that the
scattering cross section defined by (\ref{sig}) depends on the
chosen shape and size of the volume element $d{\bf v}_1^\prime$
and should not be considered as a well-defined quantity.
\vskip-0.22cm \hspace{1.3cm} \setlength{\unitlength}{0.012in}
\begin{picture}(300,165)

\multiput(85,86)(90,0){3}{\makebox(8,8)[c]{$S$}}
\put(57,35){\makebox(8,8)[c]{\bf (a)}}
\put(147,35){\makebox(8,8)[c]{\bf (b)}}
\put(237,35){\makebox(8,8)[c]{\bf (c)}}

\put(46,105){\makebox(30,8)[c]{$d{\bf v}_1^\prime$}}
\put(136,118){\makebox(30,8)[c]{$d{\bf v}_1^\prime$}}
\put(226,105){\makebox(30,8)[c]{$d{\bf v}_1^\prime$}}

\put(150,95){\oval(10,40){}} \put(240,95){\oval(40,10){}}
\put(60,95){\circle{15}}
\multiput(95.00,60.00)(90,0){3}{\circle*{1}}
\multiput(95.00,60.00)(90,0){3}{\circle*{1}}
\multiput(94.89,62.82)(90,0){3}{\circle*{1}}
\multiput(94.55,65.61)(90,0){3}{\circle*{1}}
\multiput(93.98,68.38)(90,0){3}{\circle*{1}}
\multiput(93.20,71.08)(90,0){3}{\circle*{1}}
\multiput(92.20,73.72)(90,0){3}{\circle*{1}}
\multiput(90.99,76.27)(90,0){3}{\circle*{1}}
\multiput(89.58,78.71)(90,0){3}{\circle*{1}}
\multiput(87.98,81.03)(90,0){3}{\circle*{1}}
\multiput(86.20,83.21)(90,0){3}{\circle*{1}}
\multiput(84.25,85.24)(90,0){3}{\circle*{1}}
\multiput(82.14,87.11)(90,0){3}{\circle*{1}}
\multiput(79.88,88.80)(90,0){3}{\circle*{1}}
\multiput(77.50,90.31)(90,0){3}{\circle*{1}}
\multiput(75.00,91.62)(90,0){3}{\circle*{1}}
\multiput(72.41,92.73)(90,0){3}{\circle*{1}}
\multiput(69.74,93.62)(90,0){3}{\circle*{1}}
\multiput(67.00,94.29)(90,0){3}{\circle*{1}}
\multiput(64.22,94.74)(90,0){3}{\circle*{1}}
\multiput(61.41,94.97)(90,0){3}{\circle*{1}}
\multiput(58.59,94.97)(90,0){3}{\circle*{1}}
\multiput(55.78,94.74)(90,0){3}{\circle*{1}}
\multiput(53.00,94.29)(90,0){3}{\circle*{1}}
\multiput(50.26,93.62)(90,0){3}{\circle*{1}}
\multiput(47.59,92.73)(90,0){3}{\circle*{1}}
\multiput(45.00,91.62)(90,0){3}{\circle*{1}}
\multiput(42.50,90.31)(90,0){3}{\circle*{1}}
\multiput(40.12,88.80)(90,0){3}{\circle*{1}}
\multiput(37.86,87.11)(90,0){3}{\circle*{1}}
\multiput(35.75,85.24)(90,0){3}{\circle*{1}}
\multiput(33.80,83.21)(90,0){3}{\circle*{1}}
\multiput(32.02,81.03)(90,0){3}{\circle*{1}}
\multiput(30.42,78.71)(90,0){3}{\circle*{1}}
\multiput(29.01,76.27)(90,0){3}{\circle*{1}}
\multiput(27.80,73.72)(90,0){3}{\circle*{1}}
\multiput(26.80,71.08)(90,0){3}{\circle*{1}}
\multiput(26.02,68.38)(90,0){3}{\circle*{1}}
\multiput(25.45,65.61)(90,0){3}{\circle*{1}}
\multiput(25.11,62.82)(90,0){3}{\circle*{1}}
\end{picture}


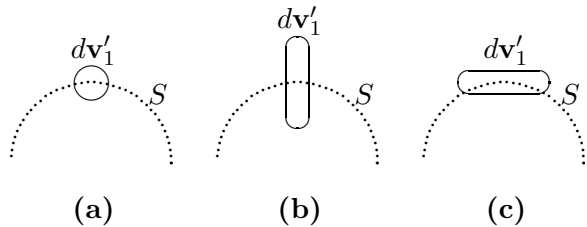
\vskip -0.5cm \centerline{Figure~1: Several possible shapes of the
velocity  element $d{\bf v}_1^\prime$.} \vskip 0.5cm

The investigation above states that many formulas in the standard
derivation, from (\ref{sig}) to (\ref{cl}), are not as meaningful
as the conventional wisdom assumes. To view the issue in a clearer
perspective, we now follow the standard approach as closely as
possible and evaluate the net change of molecular number as
directly as possible (which means the evaluation will be done
without using $\bar\sigma$).

Firstly, we are concerned with the molecules leaving, due to collisions,
a fixed volume $d{\bf r} d{\bf v}_1$ during $dt$. To formulate these
molecules, the following three steps appear to be essential.
(i) $f({\bf v}_1) d{\bf v}_1$ and $f({\bf v}_2) d{\bf v}_2$ are identified
as two colliding beams in the laboratory frame.
(ii) By switching to the center-of-mass frame, the number of collisions
is formulated as
$dt\int_\Omega u [f({\bf v}_2)d{\bf v}_2] [d{\bf r} d{\bf v}_1f({\bf v}_1)
\sigma(\Omega)d\Omega]$,
where the cross section $\sigma(\Omega)$, instead of the troublesome
$\bar \sigma$, has been employed. (iii)
Integrating the formula in the last step over ${\bf v}_2$, the number of
the molecules in $f({\bf v}_1)d{\bf r} d{\bf v}_1$
involving collisions during $dt$ is found to be
\begin{equation} \label{cl11}
dt d{\bf r} d{\bf v}_1 \int_{{\bf v}_2,\Omega} f({\bf v}_1)
f({\bf v}_2) u\sigma(\Omega) d\Omega
d{\bf v}_2.\end{equation}
Comparing (\ref{cl11}) with the second term of (\ref{cl}) seemingly suggests
that things went as smoothly as expected, but, if one wishes to further
conclude that the molecules
represented by (\ref{cl11}) are just those that leave $d{\bf r} d{\bf v}_1$
during $dt$ because of collisions, a mathematical
paradox arises sharply. Note that $dt$, $d{\bf r}$ and $d{\bf v}_1$ are
three quantities that are chosen independently. If $|{\bf v}_1 dt|$ is
much larger than $|d{\bf r}|$, where $|d{\bf r}|$
represents the length scale of $d{\bf r}$, it is easy to see that all the
molecules in $d{\bf r}d{\bf v}_1$, suffering collisions or not, will leave
the volume element during $dt$ anyway. In other words, in order for
(\ref{cl11}) to make the ``proper'' sense,
we need to assume that $|{\bf v}_1 dt|<< |d{\bf r}|$. This assumption is,
unfortunately, not more correct than the contrary.

Then, molecules that enter $d{\bf r}
d{\bf v}_1$ during $dt$ due to collisions are under investigation.
We take the following four steps to do that.
(i) $f({\bf v}_1^\prime) d{\bf v}_1^\prime$ and
$f({\bf v}_2^\prime) d{\bf v}_2^\prime$ are identified as two
colliding beams. (ii) The number of collisions is
similarly formulated in the center-of-mass frame as
\begin{equation}\label{cl2} dt d{\bf r}^\prime \int_\Omega
u [f({\bf v}_1^\prime) d{\bf v}_1^\prime][f({\bf v}_2^\prime)
 d{\bf v}_2^\prime] \sigma(\Omega)d\Omega,\end{equation}
where $d{\bf r}^\prime$ has purposely been assumed to be different
from $d{\bf r}$ for a reason that will be given soon.
(iii) It is now in order to return the laboratory frame, where
$d{\bf r}d{\bf v}_1$ is defined, and determine
what fraction of the type 1 colliding molecules expressed by (\ref{cl2})
enter $d{\bf r} d{\bf v}_1$. Surprisingly enough,
the task, if defined as that in the standard approach, cannot be
accomplished in any meaningful way. For this to be seen, follow
the standard approach and let $d{\bf r}^\prime$ be identical with $d{\bf
r}$, which reduces, with one's notice or not,
the spread issue (how molecules spread after collisions) in
the six-dimensional phase space to the one only in the
three-dimensional velocity space. Along this road, we come back to the
``old location of the maze'': for the given
$f({\bf v}_1^\prime) d{\bf v}_1^\prime$ and $f({\bf v}_2^\prime) d{\bf
v}_2^\prime$, the density of the type 1 molecules
entering $d{\bf v}_1$ after collisions depends on the shape and size of
$d{\bf v}_1$ (varying from zero to infinity). With this kind of uncertainty,
the entire formulation becomes meaningless.
(iv) If the last step were somehow completed meaningfully (which will be
 done elsewhere)
appropriate integrations should be performed to get the entire number of
 molecules entering $d{\bf r} d{\bf v}_1$ during $dt$ because of
collisions.

To get a brief and heuristic understanding of the issues raised above, let
us think about the following thought experiment. Suppose that all the
microscopic states of molecules in a gas are completely known
at $t=0$. At $t=T>0$, we measure all molecules and
find that the molecules that suffered collisions during $0<t<T$
acquire not only new velocities but
also new positions (in comparison with the `old' positions determined
by their collisionless trajectories). That is to say,
for any given time period, collisions alter
molecular velocities and molecular positions with roughly the same
`efficiency' and the two effects have to be investigated in the
six-dimensional phase space on an equal footing.

A complete discussion on involved questions and problems is much beyond
the scope of this brief paper. In some of our recent
works\cite{ref:chen1}\cite{ref:chen2}, we make more analyses and try to
introduce an alternative approach.

The work is partly supported by the fund provided by Education
Ministry, PRC.

\end{document}